# High-Level Synthesis Case Study: Implementation of a Memcached Server

Kimon Karras
Xilinx Research Labs
Dublin, Ireland
Email: kimonk@xilinx.com

Michaela Blott
Xilinx Research Labs
Dublin, Ireland
Email: mblott@xilinx.com

Kees Vissers
Xilinx Research Labs
San Jose, USA
Email: keesv@xilinx.com

*Abstract*—High-Level Synthesis (HLS) aspires to raise the level of abstraction in hardware design without sacrificing hardware efficiency. It has so far been successfully employed in signal and video processing but has found only limited use in other areas. This paper utilizes a commercial HLS tool, namely Vivado® HLS, to implement the processing of a common data center application, the Key-Value Store (KVS) application memcached, as a deeply pipelined dataflow architecture. We compared our results to a fully equivalent RTL implementation done previously in our group and found that it matches its performance, yields tangible improvements in latency (between 7-30%) and resource consumption (22% in LUTs and 35% in registers), all while requiring 3x less lines of code and 2x less development time. The implementation was validated in hardware on a Xilinx® VC709 development board, meeting timing requirements for 10Gbps line rate processing.

## I. INTRODUCTION

FPGAs have long shown great promise in accelerating computation for a multitude of applications [1], but their adoption has been hampered by a very low level programming environment which makes code hard and costly to develop as well as inflexible. This is particularly relevant in a highly dynamic environment such as data centers, where a reduction in FPGA development time is fundamental to enable operators to implement and deploy part of their volatile software stacks on FPGAs. Considerable research has flown into changing this, with C-to-RTL synthesis being one of the most popular approaches. One tool which employs this technique is Xilinx's® Vivado® HLS [2]. It uses standard C/C++ syntax with some extensions to allow for constructs typically encountered in hardware design such as bit vectors and streams.

Although HLS has so far found considerable success in the signal processing domain [3], [4], it has been limited to niche cases in other domains [5]. Many data center applications are similar in nature to network processing and as such acutely distinct from signal processing. Data structures are considerably more complex, consisting of stacks of headers, whereby each header contains multiple fields and depends in format and size on previous headers. The exact number and types of headers and/or fields in the stack is usually not known beforehand as it depends on values found within the header itself and involves keeping state during processing. In contrast, signal processing data types are more symmetric and simple, typically consisting of vectors and matrices, all being of fixed size. Computation is of a more repetitive nature while being highly intensive, requiring mathematical operations on their input at a high frequency.

This work uses a KVS application called memcached [6], as an example for a look into how HLS can be used in the context of data center applications. Memcached is well-suited to this task because its functionality covers a diverse functional set. This includes parsing headers with varying formats, realigning fields and dynamically searching for specific characters in data words, accessing external memories and ensuring transactional memory consistency.

Our results show that implementing such a complex system with Vivado® HLS is not only possible, but that the resulting design is comparable to a previously developed RTL prototype by us and described in [7], with both designs being capable of 10 Gbps line-rate processing of memcached requests, while offering significant advantages in productivity (and thus reducing Non-Recurring Engineering costs) and more succinct, easy-to-maintain source code. This is accomplished through leveraging advantages offered by HLS such as easier design space exploration, flow control and high-level programming constructs. Finally, the HLS implementation comes out ahead in resource use and latency.

The rest of this paper is structured as follows: Section II provides an overview of the efforts found in the area of high level synthesis. Section III briefly introduces memcached and the implemented architecture, while IV evaluates the HLS implementation in comparison to our RTL design in regards to abstraction level in the design entry, performance (throughput & latency), and FPGA resource requirements. Finally, Section V summarizes our findings.

## II. RELATED WORK

Conversion from higher level languages to RTL or direct gate-level synthesis has been investigated both on an academic and on a commercial front. Although early HLS efforts go back into the mid-80s [8], it is only recently that HLS is gaining broader commercial adoption mostly with companies that have a signal processing application focus ( [9], [10]). These HLS solutions include Calypto Design Systems' Catapult C [11], Forte Design System's Cynthesizer [12], Bluespec's HLS tool [13], Xilinx's® Vivado® HLS [2] and others. Most of these tools (with the exception of Bluespec) use C/C++ as their input language while Bluespec uses its proprietary BSV language which is an extension to SystemVerilog. The standard approach used by all C-based tools is to use directives to guide the implementation of the circuit. This is typically performed by inserting pragma directives in the source code which can be applied either to a function (to e.g. tell the compiler to pipeline that function) or to a specific loop (to e.g. unroll or flatten it) or even to a specific variable (to e.g. set if this variable is to be implemented as a memory). The tool then synthesizes the code and produces RTL which can be simulated and runs through logic synthesis as required. In some tools, C simulation can be performed to achieve functional correctness of the design before proceeding with the lengthy design implementation flow.

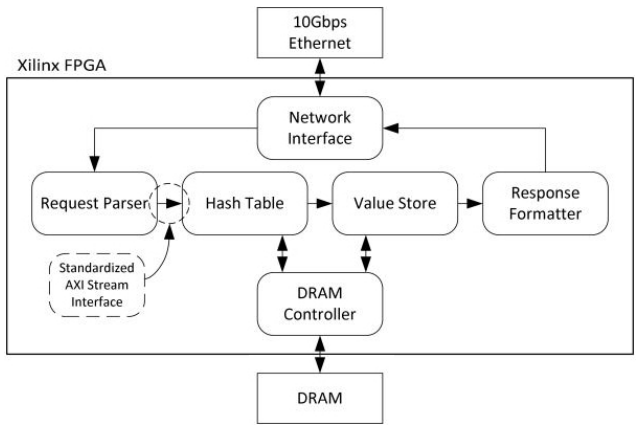

Fig. 1: Memcached data flow architecture

| S | E | T | space | KEY <LK BYTES> | | space |
|---|---|---|---|---|---|---|
| Flags (ASCII encoded) | | | space | Expiration (ASCII encoded) | | space |
| ValueLength (ASCII encoded) | | | (0d) | (0a) | VALUE <LV BYTES> | (0d) |
| (0a) | | | | | | |

Fig. 2: Memcached ASCII SET request format

Independently of the flavor of the language used, most of these efforts have in common a tendency to address the same problem, namely signal and video processing applications, in which the same operation has to be applied to consecutive samples or pixels. These applications are repetitive in nature and thus highly amenable to parallelism which in turn enables the use of loop manipulation (unrolling, etc.) to quickly investigate different architectural alternatives. There is significant literature available on these efforts and commercial tools in this area have indeed managed to find considerable adoption. This notwithstanding, high-level synthesis tools are almost completely absent from any other application domain.

Simultaneously, domain Specific Languages like SDNet [14] have managed to raise the level of abstraction in FPGA programming by offering a very software like approach to specific problems and hiding hardware-related details from the developers. While they focus on a specific set of application and thus have a narrow scope, they provide a higher abstraction level than typical HLS approaches.

Finally, in regards to other implementations of memcached on FPGAs, we would like to refer to Derek Chiou's work [15], who leveraged hybrid architectures, deploying a mixture of CPUs and FPGA fabric, rather than a standalone FPGA implementation. A key focus of their work included a profiling tool, which identified the most frequently executed traces for acceleration within the FPGA. Similarly, another effort was conducted at HP [16], advocating a hybrid architecture. Both bodies of work did not experiment with HLS tools as design flow.

III. MEMCACHED PIPELINE ARCHITECTURE

Key-Value Stores such as memcached have become commonplace as middleware in data centers where they are used as a caching tier between web servers and databases to overcome scalability issues. Their basic functionality is that of a networked associative memory. Typical operations include a GET command which retrieves the value associated with a key, as well as a SET command which writes a key-value pair into the store. Memcached uses mainly two protocols, a binary and an ASCII variant. Each protocol includes about 15 commands which can be broadly categorized into storage commands (SET, ADD, APPEND, etc.), which store data related to a corresponding key, and retrieval commands (GET, GETS, CAS, etc.), which read previously stored data. Our implementation focuses on a selection of core commands, namely SET, GET, DELETE and FLUSH, thereby demonstrating all basic functionality.

The binary protocol uses a common message format for all request and response types. The message consists of a fixed length header, which contains all the information that is needed to extract all required fields from the payload of the packet. This is akin to typical networking protocols such as Ethernet where the location and size of all fields in the byte stream are predetermined.

In the ASCII protocol the message formats vary profoundly and ranges from very simple to vastly complex. An example of the former is the SET response, which is a fixed message returned by the memcached server in case of a successful SET operation. In contrast, a SET request, shown in Figure 2, consists of a small fixed part (the first four bytes) followed by five fields arranged in a completely unpredictable pattern. All fields are of variable and unknown length with some of them being delimited with 0x20 and some with 0x0D0A. Furthermore, as a result the offset to the next field is unknown and can only be determined after the previous field is extracted. This means that when trying to extract for example the key, all bytes of all data words have to be scanned until the character 0x20 (or 0x0D0A) is found and this has to be repeated for all other fields. This search process is further complicated by the fact, that up to three separate fields can be contained in one data word (64bit in our case) of an incoming stream and must be processed within a clock cycle to achieve full line-rate throughput. This constitutes the key challenge in parsing the ASCII-based protocols.

In order to accelerate memcached using an FPGA, we have transformed the application to fit a streaming data flow architecture as shown in Figure 1. This architecture is described in much greater detail in [7]. Packets are received and transmitted through a network interface that includes basic Ethernet, UDP and TCP processing. Only the memcached part of the packet is then passed through a 4-stage pipeline. First the requests are fed into the parser module, which detects the protocol used and transforms the request into a common internal pipeline format. This is then passed to the hash table, which performs a look-up and retrieves the memory address to be accessed by the value store. This is the next stage, which performs the actual memory access to write or fetch the value from the memory. Finally, the response formatter module converts the internal pipeline format into the appropriate memcached protocol response. The internal structure of the four stages is further detailed in the following paragraphs.

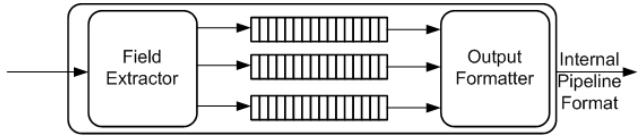

Fig. 3: Memcached binary parser architecture

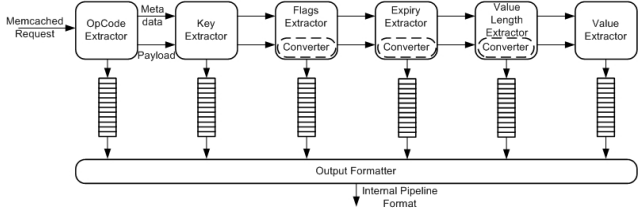

Fig. 4: Memcached ASCII parser architecture

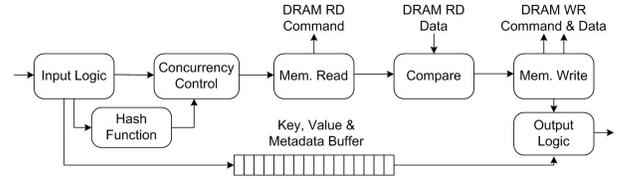

Fig. 5: Memcached hash table architecture

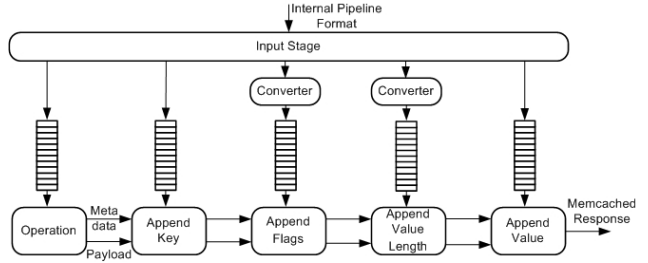

Fig. 6: Memcached ASCII response architecture

*A. Request parser*

Parsing the memcached binary and ASCII protocols are two fundamentally different tasks. In the former all the field locations and length are known beforehand. Thus, the simple parser, show in Figure 3, consists of two finite state machines (FSM) joined together by a set of buffers. The first state machine, which we call field extractor, reads the incoming packets and extracts from them the key, the value and a set of metadata, which includes the key and value length and additional information required for further processing, and writes these into the intermediate buffers. From there they are read by the second FSM, which then formats and forwards them onto the next stage in the pipeline.

The ASCII parser is more complex, as can be seen in Figure 4, however the basic architecture resembles the binary parser in that it consists of a 6-stage pipeline, with each stage handling the extraction of exactly one field similar to the field extractor of the binary parser. The extracted fields are then passed to an output formatter, which creates the internal pipeline format required. In addition, some of these stages include a converter to handle necessary ASCII to binary conversion for fields that require further computation such as expiration time. The extracted fields are then collected by the output formatter from the intermediate buffers and produces the required output format. The code for the extractor stages is identical whereby some of them contain a converter as an additional sub-module.

*B. Hash table*

The hash table is divided into 7 distinct pipeline stages as illustrated in Figure 5. First an input logic splits the header information from the payload, which is buffered separately, and the output logic at the end of the pipeline inverses the process. Between them, the hash function unit computes a Bob Jenkins hash over the key, before the packet enters the concurrency control. The latter unit ensures memory consistency between transactions and is detailed further in the evaluation below. The memory read unit fetches the contents of the hash table at the location given by the hash which is then compared with the currently processed key. Depending on whether this is a hit or a miss, and the type of operation, the hash table entry is updated in the memory write unit. Further detail on the hash table can be found in [17].

*C. Value store*

The value store implementation consists of 5 blocks. Similar to all other modules, it contains an input logic, which decodes the operation, pushes unnecessary data into a buffer and moves the rest either into a read or a write pipeline stage. The read unit issues read requests to retrieve a value from DRAM while the write unit inserts or updates a value in memory. At the output of the module, an output logic unit can be found, which merges all data and reformats the output.

*D. Response Formatter*

The architecture of the ASCII response formatter is shown in Figure 6 whereas the binary response formatter is a simplified subset of this, as all field sizes and offsets are predetermined. In the input stage of the response formatter, the data is split into five parts representing distinct sections of the response packet. Flags and value length have to be converter back into ASCII before all fields are buffered in an intermediate set of FIFOs. The output formatting is divided up into five separate stages, whereby each stage is responsible for one section of the packet. Similarly to the ASCII parser, the additional stages for packet assembly are necessitated as the length of the ASCII fields and with that the offsets to subsequent fields are not known beforehand. Therefore. a single FSM would become far more complex.

## IV. IMPLEMENTING A MEMCACHED SERVER USING VIVADO® HLS

This section highlights some of the core challenges encountered in implementing an FPGA memcached server using Vivado® HLS, elaborates on the results of our implementation and compares it with the previously completed RTL one. We show that significant improvements in the abstraction of the design can be achieved while delivering equivalent performance in throughput and bringing connsiderable resource and latency advantages.

### A. HLS Benefits: Design Abstraction

*1) FIFO/Memory Access and Flow Control:* One of the major sources of errors in the RTL implementation were related to flow control. Vivado® HLS eliminates this hurdle by abstracting flow control away from the user. The tool takes care of any FIFO instantiations, read & write interface signalling, addressing, and flow control (full and empty signals) which indicate validity of data words and ability to forward as the code snippet below shows. In the RTL equivalent, every state transition would be conditioned by forward and backward flow control signals. This eliminated around 80% of errors encountered in the RTL design and made the source code leaner and simpler.

```
if (!inData.empty()) {
 inData.read(inputWord);
 switch(countState) {
  case IDLE:
       counter = 1;
       countState = COUNT;
       break;
  case COUNT:
       counter++;
       if (inputWord.last && !lengthOut.full()) {
        lengthOut.write(counter);
        countState = IDLE;
       }
       break;
 }
}
```

*2) High-Level Programming Constructs:* A key advantages of using higher level programming languages for hardware design is the additional constructs that the designer can leverage. A prime example of this is found in the hash table's concurrency control stage. It consists of a look-up table which stores all concurrently executing SET or DELETE operations. For each incoming GET, a look-up into this table is carried out to determine any possible access conflicts. Furthermore, it is necessary to add and remove elements from the table as SET and DELETE operations enter and leave the critical section of the hash table. The exit point of the critical section is the memory write stage. As such, this data structure is queue-like with a parallel look-up functionality.

In HLS, we can describe this table as a class (see code below) thereby providing the programmer with all typical advantages associated with object-oriented programming.

```
class concurrencyFilter {
  private:
    uint8_t           wrPtr;
    uint8_t           rdPtr;
    filterEntry fEntries[noOfEntries];
  public:
    concurrencyFilter();
    bool push(filterEntry newElement);
    bool pop();
    bool compare(filterEntry compElement);
};
```

The class contains an array (fEntries) where the entries itself are stored. A write and a read pointer (wrPtr, rdPtr) is used to keep track of where to write and read in the array and as such implementing the queue functionality. Furthermore, three different methods are declared: compare is the method that carries out the actual look-up within the array, while push and pop allow adding and removing of entries from the table as commands enter and leave the critical section.

TABLE I: HLS synthesis timing results for various clock constraints

| Clock Constraint | Reported Clock Period | Latency |
|---|---|---|
| 10 | 8.21 | 3 |
| 6.6 | 5 | 5 |
| 5 | 3.77 | 8 |
| 3.3 | 2.5 | 9 |
| 1.6 | 1.54 | 13 |

*3) Module Pipelining:* A further benefit of HLS is automatic pipelining of modules in order to meet timing requirements. This is particularly prevalent in the ASCII to binary conversion which we discussed in Section III-A and illustrated in the code snippet below (Binary to ASCII conversion is equally concise). HLS automatically pipelines the resulting logic by analyzing the design requirements (clock period, any latency limitations, etc.). In our application, latency is for instance less critical, whereas our clock frequency is bound by the 10 Gbps Ethernet clock rate to 156 MHz. Thus, HLS automatically produces a circuit with a latency of 5 cycles which meets the timing requirement. Changing the target clock period can affect the latency of the generated circuit as shown on table I as HLS increases the pipeline depth accordingly to meet the timing requirements.

```
for (unsigned short int i=0;i<10;++i) {
 subArray((4*(i+1))-1, i*4) =
 inData((8*(i+1))-1, i*8) - 48;
 sum += subArray((4*(i+1))-1, i*4)*(10*i);
}
```

*4) Design Space Exploration:* Another fundamental advantage of high abstraction levels is the simplicity of design space exploration to explore numerous alternative implementations and evaluate them in regards to key metrics such as performance, latency, and resource usage. This has proven to be particularly useful in the context of parsing an ASCII-based protocol. One of the main issues in this context is that it contains sequences of fields of unknown length, which are delimited by a specific character sequence. Thus all bytes of each data word have to be searched for these delimiter characters over multiple cycles. Additionally, the field to be extracted might not begin at the first byte of a data word. Thus an offset from the beginning of the first data word has to be taken into account. This can be seen in Figure 2. The code snippet below shows how such a search can be implemented optimally in HLS by using a loop to parse and compare all the data word bytes. First we perform a shift operation to eliminate the offset. Then a for-loop iterates through all bytes in reverse order and looks for the delimiting character. When found, the location is stored. The loop has to have fixed bounds in order for the synthesizer to be able to unroll it completely.

```
shData.range(((8-offset)*8)-1, 0) =
inData.range(63, (offset*8));
for (unsigned short int i=8;i>0;--i) {
 if (shData.range((8*(i+1))-1, 8*i)==0x20)
  startLoc = i;
}
```

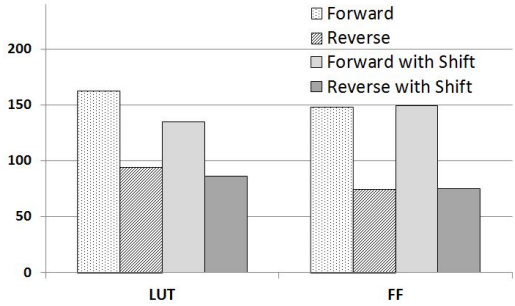

Fig. 7: Resource use for the 4 code variants

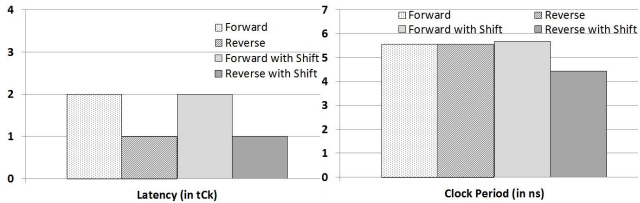

Fig. 8: Latency and clock period for the 4 code variants

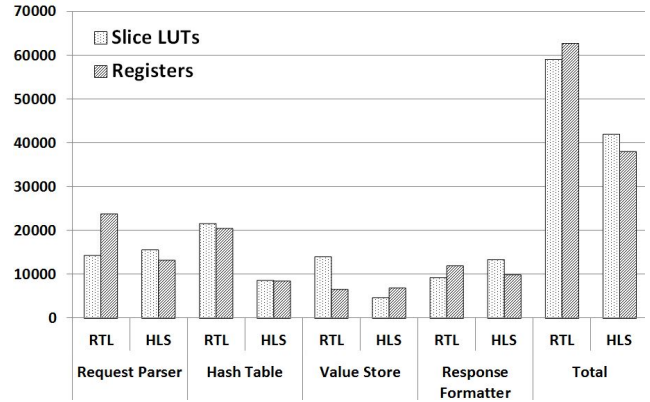

Fig. 9: LUT and Register in the memcached prototype

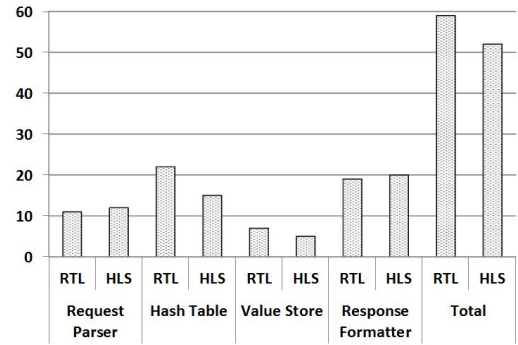

Fig. 10: BRAM use in the memcached prototype

It took a number of iterations over multiple variants to find the optimal one. We analyzed three additional variants of the aforementioned approach. The first variant searches forward in the data word without shifting the data word before. The second variant performs a backwards search without shifting, while the third one shifts the input data first and then searches forward. Figures 7 and 8 show the results for these four variants. It is immediately evident that reverse-searching yields better results in comparison to the forwarding-searching ones, which is used in the forward-looking variants as Figure 8 shows. The reason behind this is that the forward-looking variants require a break statement, which has to be included in the loop when the first character was found and which results in an additional comparison step to determine the required match. This is required to ensure that there are no more identical delimiting characters not belonging to this field in the same data word after the search field ends.

### B. Implementation Results

An HLS implementation is only attractive if the resource, throughput and latency requirements of the application can be met. In this section, we compare the HLS implementation with our RTL implementation of the exact same functionality.

*1) Resource Use:* From a resource use perspective, the HLS memcached implementation is clearly superior to the RTL one as shown in Figures 9 and 10, which illustrate the resource requirements of the four key pipeline stages. All of the modules were synthesized for a Xilinx® XC7VX690T FPGA using Vivado® 2013.4. The results for the total system refer to the synthesis of the entire pipeline and thus are smaller than the sum of the four components, as the tool is able to further optimize and combine resource use.

An area where HLS is consistently superior to the RTL implementation is the register use. This is due to HLS being able to optimize intermediate buffering of data words inside state machines and thus trim resources significantly. This is particularly evident in the context of the hash table, which buffers the 512-bit wide data words coming from the memory internally for processing. Furthermore, as HLS state machines have explicitly described fewer states and state transitions than in RTL (as flow control is handled behind the scenes), the overall length of the data path is reduced and again many Flip Flops can be saved. This also impacts the total latency of the pipeline as can be seen further below.

Furthermore, improved resource consumption of the HLS implementation is to be attributed to slight differences in the implementations, which were enabled by HLS' pipelining capabilities. This is exemplified by the implementation of the Bob Jenkins [18] hash algorithm. This algorithm includes a series of interdependent arithmetic and logical operations which have to be performed over multiple clock cycles to meet the timing requirements. Creating such a pipelined state machine in RTL is considerably complex and thus in the RTL version, 8 parallel instances of the hash function are used to meet the throughput requirement. HLS does this pipelining automatically and thus only one instance of the hash function is required, which leads to a resource use reduction of almost 70% for the hash function as illustrated in table II (the RTL numbers are for all 8 instances of the module).

TABLE II: Bob Jenkins hash function resource use

|     | LUTs | Registers |
| --- | --- | --- |
| HLS | 2496 | 799 |
| RTL | 7408 | 2352 |

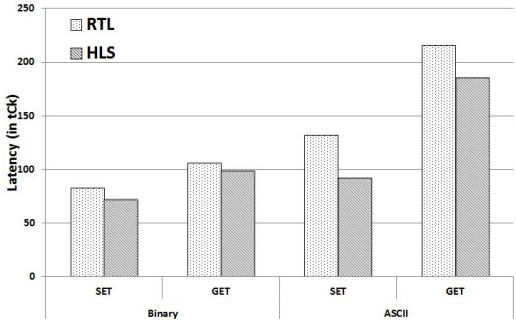

Fig. 11: Memcached pipeline latency

TABLE III: Lines of code for each memcached pipeline stage

|     | Request Parser | Hash Table | Value Store | Response Formatter | Total |
| --- | --- | --- | --- | --- | --- |
| RTL | 3259 | 2619 | 3196 | 2283 | 11359 |
| HLS | 1493 | 964 | 1060 | 552 | 4069 |

*2) Design Performance:* Performance-wise both designs are capable of achieving 10 Gbps line-rate performance, which is a 9x improvemennt in performance over the best reported results for x86 processor for small packet sizes as reported [7]. in Achieving timing closure for the HLS design required special attention since the use of Vivado® HLS added one additional level of abstraction in the design. High-Level Synthesis provides the user with a frequency estimate after synthesizing the C++ source code. This estimate proved to be overly pessimistic in our design. Synthesizing the generated code with Vivado® synthesis eliminated most of the timing issues that were reported by HLS synthesis. This is a limitation of Vivado® HLS which does not perform the broad logic optimizations of a full-fledged synthesis tool and thus lacks the latter's precision. The critical path of our design after logic synthesis was located in the hash table's concurrency control and consisted of 13 layers of logic, which caused the design to barely meet timing requirements. The underlying cause was the inlining of the concurrencyFilter class (described in IV-A2) methods. It was possible to identify the root cause by correlating signal names in the timing report with the HLS source code.

A latency comparison between various commands is shown in Figure 11, clearly highlighting that the HLS implementation improves latency overall. This is to be attributed to HLS' capability to optimize the pipeline depths of specific modules (e.g. the Binary-to-ASCII converters) effortlessly, balancing logic over the minimum required number of pipeline stages, something very time-consuming in RTL design.

Finally, the main differentiator between the two implementations was the complexity of the resulting code and the related development effort. Table III captures this by comparing the lines of code for each module. HLS code is between 25-50% of the RTL version of the same module, resulting in a 64% overall reduction for the entire design. Development effort is harder to quantify and subject to many other factors such as programmer's skill levels. However roughly speaking, we have seen a 50% reduction in development time in comparison to RTL.

## V. Conclusions

This work presented the implementation of a memcached server with Vivado® HLS and used this implementation as a vantage point for evaluating the suitability of Vivado® HLS for large data center applications. The HLS implementation achieved equivalent throughput (10 Gbps line rate performance) while requiring less resources than the respective RTL design. The results show that Vivado® HLS can produce code of comparable efficiency to hand-written RTL, while at the same time significantly raising the level of abstraction and thus increasing productivity. This is accomplished through the complete abstraction of flow control, easy design exploration and the use of high-level code constructs, without sacrificing designer control over the architecture.

This shows that HLS is able to tackle large, complex designs in the data center domain with great efficacy. The next step in our investigation is to use OpenCL to integrate the existing HLS implementation with a host processor.

## References

[1] Scott Sirowy et al. *Where is the Beef? Why FPGAs Are So Fast.* Microsoft Research TechReport, 2008.

[2] Xilinx Inc. *Vivado Design Suite User Guide: High-Level Synthesis.*

[3] K. Vissers, et al. *Building real-time HDTV applications in FPGAs using processors, AXI interfaces and high level synthesis tools.* In DATE'11, pages 1–3, 2011.

[4] J. Cong, et al. *High-Level Synthesis for FPGAs: From Prototyping to Deployment.* Computer-Aided Design of Integrated Circuits and Systems, IEEE Transactions on, 30(4):473–491, 2011.

[5] Nathaniel Jachimiec et al. *High-Level Synthesis Tool Delivers Optimized Packet Engine Design.* XCell Journal, 2:14–18, 2012.

[6] Online. *http://memcached.org/.*

[7] Michaela Blott, et al. *Achieving 10Gbps Line-rate Key-value Stores with FPGAs.* HotCloud'13, November 2013.

[8] F.F. Yassa, et al. *A silicon compiler for digital signal processing: Methodology, implementation, and applications.* Proceedings of the IEEE, 75(9):1272–1282, 1987.

[9] J.G. Mena, et al. *High level synthesis of a Front End filter and DSP engine for analog to digital conversion - a case study.* In VTS'10, pages 252–252, 2010.

[10] E. Torbey et al. *Implementation and trade-offs of a DCT architecture using high-level synthesis.* In Proceedings of the11th Annual IEEE International ASIC Conference, pages 193–197, 1998.

[11] Calypto Design Systems. *Catapult: Product Family Overview.* http://calypto.com/en/products/catapult/overview, 2013.

[12] John Sanguinetti et al. *High-Level Modeling and Hardware Implementation with General-Purpose Languages and High-level Synthesis.* 9th IEEE/DATC Electronic Design Processes Workshop, 2002.

[13] Bluespec. *High-Level Synthesis Tools.* http://www.bluespec.com/high-level-synthesis-tools.html, 2013.

[14] Gordon Brebner et al. *High Speed Packet Processing using Reconfigurable Computing.* IEEE Micro, 2013.

[15] Maysam Lavasani, et al. *An FPGA-based In-line Accelerator for Memcached.* IEEE Computer Architecture Letters, 99(RapidPosts):1, 2013.

[16] Kevin Lim, et al. *Thin Servers with Smart Pipes: Designing SoC Accelerators for Memcached.* SIGARCH Comput. Archit. News, 41(3):36–47, June 2013.

[17] Zsolt Istvan, et al. *A Flexible Hash Table Design for 10Gbps Key-Value Stores on FPGAs.* In 23rd International Conference on Field Programmable Logic and Applications (FPL'13), 2013.

[18] Bob Jenkins. *Function for Producing 32-bit Hashes for Hash Table Lookup.* http://burtleburtle.net/bob/c/lookup3.c, 2006.